\newif\ifdraft
\draftfalse   
\newif\ifanon
\anonfalse

\documentclass{article}

\usepackage[preprint]{neurips_2024}

\usepackage[scaled]{beramono}
\usepackage[T1]{fontenc}
\usepackage{dafny}

\usepackage{microtype}
\usepackage{graphicx}
\usepackage{subfigure}
\usepackage{booktabs} 
\usepackage{multirow}
\usepackage{quiver}

\usepackage{hyperref}

\usepackage{amssymb}
\usepackage{pifont}

\usepackage{listings}

\usepackage{algorithm, algorithmic}
\usepackage{wrapfig}

\usepackage{amsmath}
\usepackage{amssymb}
\usepackage{mathtools}
\usepackage{amsthm}

\usepackage[capitalize,noabbrev]{cleveref}

\theoremstyle{plain}
\newtheorem{theorem}{Theorem}[section]

\newtheorem{lemma}[theorem]{Lemma}

\theoremstyle{definition}

\theoremstyle{remark}

\usepackage[textsize=tiny]{todonotes}

\title{VerMCTS: Synthesizing Multi-Step Programs using\\ a Verifier, a Large Language Model, and Tree Search}

\author{
David Brandfonbrener\thanks{Kempner Institute at Harvard University, $\ ^\ddagger$ Harvard University, $\ ^\dagger$ TU M\"unchen, $\ ^\mathsection$ Northeastern University, $\ ^\mathparagraph$ Million.js, $\ ^\|$ University of Alabama at Birmingham, $\ ^{\dagger\dagger}$ University of Oxford\\Correspondence to \texttt{namin@seas.harvard.edu}}
\And
Simon Henniger$^\dagger$
\And 
Sibi Raja$^\ddagger$
\And 
Tarun Prasad$^\ddagger$
\And
Chloe Loughridge$^\ddagger$
\And 
Federico Cassano$^\mathsection$
\And 
Sabrina Ruixin Hu$^\ddagger$
\And 
Jianang Yang$^\mathparagraph$
\And 
William E. Byrd$^\|$
\And 
Robert Zinkov$^{\dagger\dagger}$
\And 
Nada Amin$^\ddagger$
}

\begin{document}

\maketitle

\begin{abstract}
Large Language Models (LLMs) can generate useful code, but often the code they generate cannot be trusted to be sound.  
In this paper, we present VerMCTS, an approach to begin to resolve this issue by generating verified programs in Dafny and Coq. VerMCTS uses a logical verifier in concert with an LLM to guide a modified Monte Carlo Tree Search (MCTS).
This approach leverages the verifier to gain intermediate feedback inside the search algorithm by checking partial programs at each step to estimate an upper bound on the value function.
To measure the performance of VerMCTS, we develop a new suite of multi-step verified programming problems in Dafny and Coq.
In terms of pass@$T$, a new metric which computes the pass rate given a budget of $T$ tokens sampled from the LLM, VerMCTS leads to more than a 30\% absolute increase in average pass@5000 across the suite over repeated sampling from the base language model.
\ifanon 
\fi
\unless\ifanon Our code and benchmarks are available at \tiny{\url{https://github.com/namin/llm-verified-with-monte-carlo-tree-search}}.\fi
\end{abstract}

\section{Introduction}

Large Language Models (LLMs) are increasingly used for generating code, but the code needs to be inspected and possibly re-generated if it doesn't satisfy the user \citep{zhong2023study}.
We propose to partially shift the burden of checking code, from the user to the LLM, by generating code in a verification-aware programming language like Dafny or Coq, prompting for specifications and proofs of correctness in addition to code that can then be formally verified.
In such a system, the user can focus their attention on the specifications, and less on the code and proofs with the assurance that the generated output has passed the verifier.
Our approach couples imprecise generative reasoning from an LLM with logical reasoning from a program verifier.
The LLM contributes fruitful suggestions and the verifier ensures soundness.

As a motivating example, consider this prompt:
{\it In Dafny, write an ADT for arithmetic expressions comprising constants, variables, and binary additions. Then write an evaluator taking an expression and an environment (a function that takes a variable name and returns a number) and returning the number resulting from evaluation. Then write an optimizer taking an expression and returning an expression with all additions by 0 removed. Then prove that the optimizer preserves the semantics as defined by the evaluation function.}

There are many types of errors that an LLM could make, including issues of syntax (e.g., how to define a function in Dafny), totality (e.g., handling unbound variables), semantics (e.g., is the optimizer invariant in the values of variables), and optimality (e.g., is the optimizer removing all additions by 0).

\begin{figure*}
    \centering
  \includegraphics[width=0.8\textwidth]{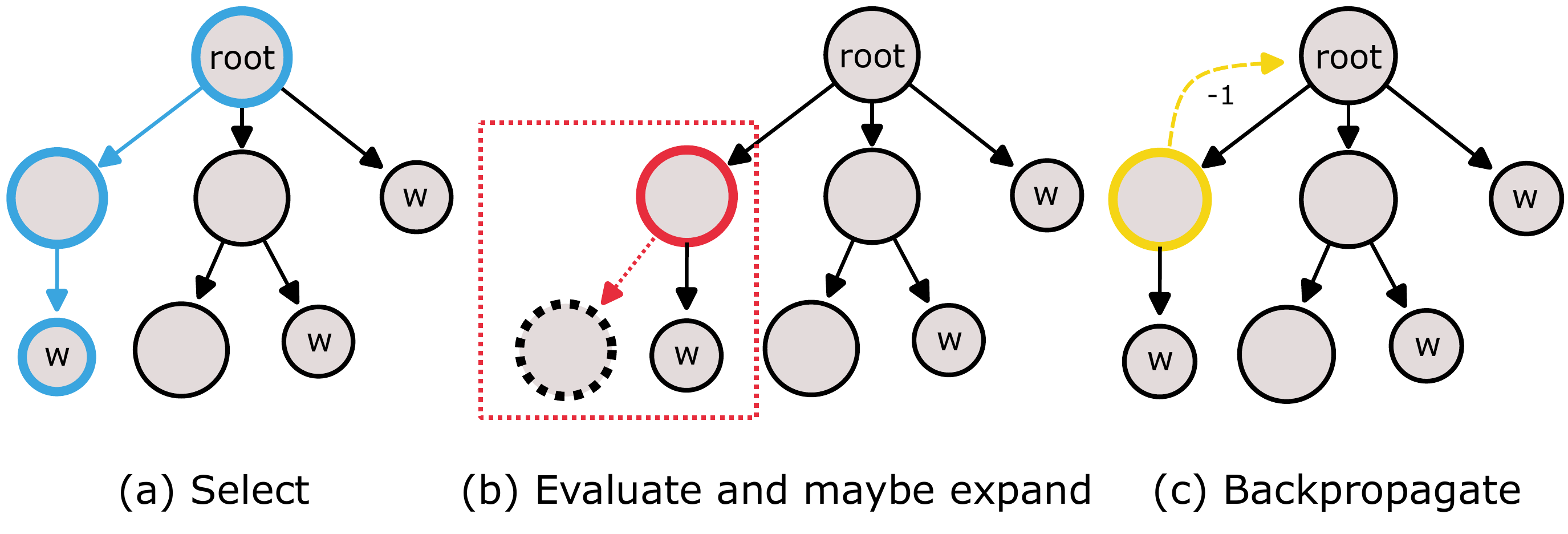}
  \vspace{-0.2cm}
  \caption{
  Overview of VerMCTS. The search tree is visualized with ``widen'' nodes marked with $w$. (a) A leaf node is selected as in standard MCTS. (b) The selected node is evaluated and maybe expanded. If the selected node is a widen node, then it's parent is selected and maybe expanded (i.e. made wider). See Figure \ref{fig:expand} for a detailed description. (c) Once we have a value and maybe from the evaluate and maybe expand algorithm, we backpropagate that value up the tree. This figure illustrates the special case where we observed a failure, so no node is added and the score is -1.}
  \label{fig:overview}
\end{figure*}

\begin{wrapfigure}[19]{r}{0.4\textwidth}
    \vspace{-0.5cm}
    \centering
    \includegraphics[width=0.4\textwidth]{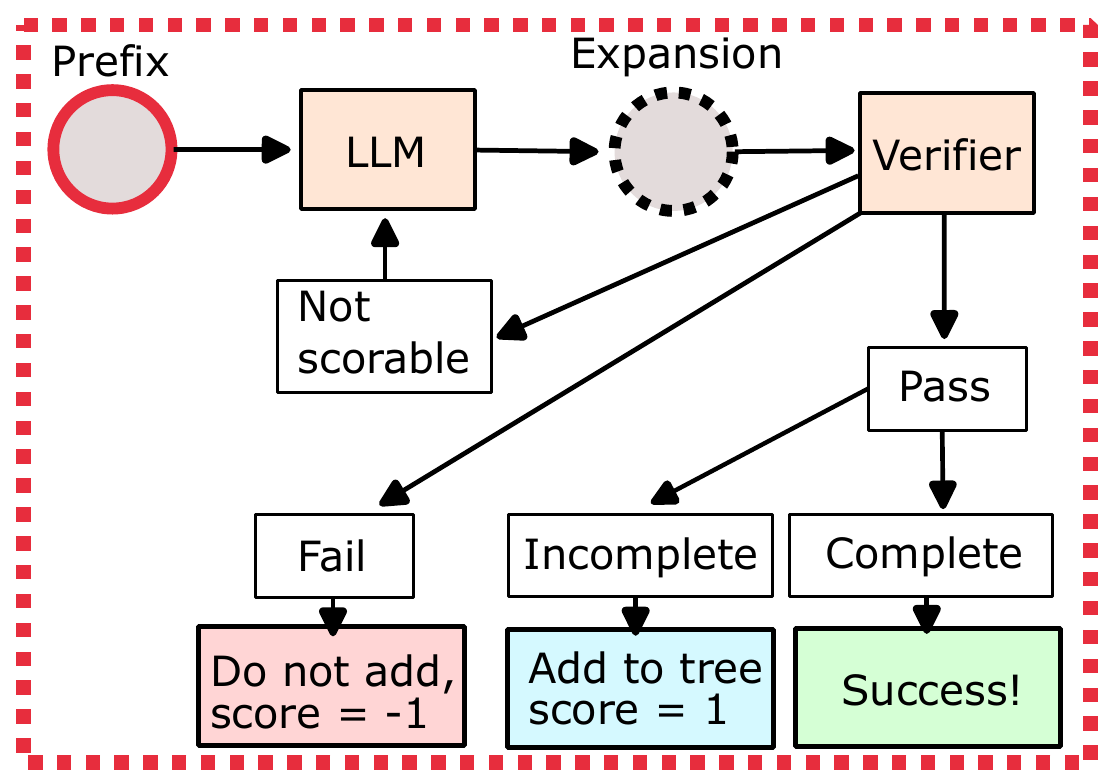}
    \vspace{-0.5cm}
    \caption{Evaluate and maybe expand. Given a prefix, we query the LLM for expansions until the verifier is able to return a score. If that score is a failure, we do not add the node to the tree, but update the parent with a value of -1. If the score is pass, then we check if the program is complete. If incomplete, we add the expansion to the tree with a score of 1. If complete, we have found a successful program to return. }
    \label{fig:expand}
\end{wrapfigure}

To aid a language model to tackle this task, we introduce VerMCTS, an algorithm that combines a verifier and tree search with a language model to synthesize verified programs. An overview of the algorithm is described in \Cref{fig:overview} and \Cref{fig:expand} and the details are presented in \Cref{sec:mcts}. VerMCTS creates a search tree with progressive widening so it is capable of handling large action spaces defined by lines of code. Within this search tree both expansion and evaluation are guided by the verifier which acts as a computationally cheap (relative to the LLM) upper bound on the value function in the code synthesis MDP, as we show in \Cref{sec:mcts}.

To evaluate VerMCTS we introduce a suite of introduce a suite of 15 challenge problems (9 in Dafny and 6 in Coq). This suite probes essential skills needed for general verified programming like constructing algebraic data types, defining functions, and writing inductive proofs. 

On this suite of problems we compare VerMCTS with several baselines including repeated sampling of full programs from the base model, an advanced prompting technique that uses access to the error messages generated by the verifier called Reflexion \citep{shinn2023reflexion}, and a traditional version of MCTS. We quantfy performance in terms of pass@$T$, the pass rate within a budget of $ T $ tokens. VerMCTS outperforms the baselines substantially, leading to a 30\% absolute average performance improvement over repeated sampling from the base model. Note this repeated sampling is a strong baseline, similar to a pass@$k$ evaluation often used as a skyline in program generation. Moreover, for several problems VerMCTS solves problems that are not solved at all by other methods within the given budget.

To summarize, our main contributions are as follows:
\begin{itemize}
    \item We introduce the VerMCTS algorithm to combine an LLM with a formal verifier and a tree search algorithm in a principled manner.
    \item We introduce a suite of challenge problems in Dafny and Coq that test key capabilities needed for verified programming.
    \item We conduct a rigorous analysis of VerMCTS and baselines on our suite of problems, finding substantial improvements when using VerMCTS.
\end{itemize}

\section{Method: VerMCTS}\label{sec:mcts}\label{sec:vllm-mcts}

Our main contribution is to define a search algorithm inspired by MCTS that leverages a verifier and LLM to search for verified programs. We call this method \emph{VerMCTS}.
In this section, we first present the Markov Decision Process that we consider as the environment for verified program synthesis and then present VerMCTS in detail. 
VerMCTS is a variant of traditional MCTS that incorporates the LLM as a prior to generate candidates and the verifier as a heuristic to evaluate partial programs.

\subsection{MDP for verified program synthesis}

We formulate our multi-step verified synthesis problem as a Markov Decision Process (MDP) $\mathcal{M} \coloneqq (\mathcal{S}, \mathcal{A}, T, r, H)$ defined by the LLM and the verifier. Here, $\mathcal{S}$ refers to the state space, $A$ refers to the action space, $T \colon \mathcal{S} \times \mathcal{A} \rightarrow \mathcal{S}$ refers to the 
 (deterministic) transition dynamics of the environment, $r: \mathcal{S}\to \mathbb{R}$ refers to the reward function, and $ H $ is the finite horizon (i.e. a limit on the number of transitions). Defining the MDP just consists of defining these four objects. The state, action, transition dynamics, and reward are defined as follows:
\begin{itemize}
    \item Each state $ s \in \mathcal{S}$ is a string consisting of the initial user prompt and a partial program.
    \item Actions $ a \in \mathcal{A}$ are strings that represent a unit of a program. In Dafny each line is an action. In Coq each ``command'' (ending with a dot `.') is an action. We also limit the number of tokens in an action.
    \item The transition dynamics are just defined by string concatentation: $ T(s,a) = s + a$.
    \item The reward function $ r$ is defined by the verifier for a given verified programming language and is only defined on complete programs. This terminal reward is 1 if the complete program is accepted and -1 if it is rejected. The reward is 0 for incomplete programs. 
\end{itemize}
With this simple MDP in place, we can define our search algorithm for finding verified programs.

\subsection{VerMCTS}

Given this MDP with finite actions and deterministic dynamics, it would be possible to run standard MCTS to learn a stochastic policy, but the action space is much too large for this to be practical. 
Instead, we build a search algorithm inspired by MCTS that can leverage the LLM as a prior for program synthesis and the verifier to evaluate partial programs. Both components are key for a successful search in this large space.

Standard MCTS consists of four steps: select, expand, evaluate, and backpropagate. Our algorithm leaves the select and backpropagate steps essentially unchanged. We modify and combine the expand and evaluate steps to leverage the power of the LLM and the verifier in tandem. Our full algorithm is illustrated in \cref{fig:overview}. In this section we first discuss progressive widening and then go through each step of VerMCTS in turn.

\paragraph{Progressive widening.} 
The naive approach to expansion using the LLM is to sample $k $ program chunks from the LLM conditioned on the current partial program and to commit to these $ k $ chunks as the only children of the node being expanded. However, this has two issues. First, if $ k$ is too small and none of the chunks can yield successful completions, then the search is stuck. But second, if we make $ k $ too large then we waste computation to sample $ k $ completions even on easy parts of the program.

To allow for potentially infinite width while still efficiently conducting deep searches, we adapt an idea from classical work on MCTS to progressivly widen nodes in the tree \citep{Chaslot2008ProgressiveSF, Coutoux2011ContinuousUC}. In that work, the number of children available at a given node scales explicitly with the number of visits. In our setting since adding a child node requires an expensive call to the LLM, we instead opt to add a ``widen'' child to each node that is assigned 0 value and can be selected via the selection procedure described below. This allows the scoring mechanism to prioritize when to expand a node by essentially setting a prior that unexplored branches have 0 value. When the widen node $ w $ with parent $ s $ is selected, instead of adding a child to $ w $, we add a child to $ s$ (i.e. add a sibling to $ w$). In this way, the tree can grow wider over the search process.

\paragraph{Selection: priors and UCT.} We use a standard MCTS selection step, but we set a prior for the UCT (upper confidence bound for trees) bonus as in PUCT \citep{Rosin2011MultiarmedBW, Silver2016MasteringTG}. We choose to let the prior $ p = 1.0$ for standard nodes and let $ p = p_{widen} < 1.0$ for widen nodes be a hyperparameter that we tune. This basic heuristic gives the model a preference to select the standard nodes which encourages deeper search trees while still allowing for potentially infinite width if needed. With this choice, the score of a node $ s $ is:
\begin{align}
\text{score}(s) = p_s \cdot c_\textit{UCT} \sqrt{\frac{\log{N_\textit{parent}}}{N_s}} + \frac{\sum_{i=1}^N v_i}{N_s}
\end{align}

\begin{wrapfigure}[18]{R}{0.45\textwidth}
\begin{minipage}{0.45\textwidth}
\vspace{-0.8cm}
\begin{algorithm}[H]
\begin{algorithmic}[1]
   \STATE {\bfseries Input:} string $s$, depth limit $ L$ \\
   \texttt{LLM}: string $ \to $ completion\\
   \texttt{Verifier}: string $ \to \{-1, 0, +1\}$
   \STATE {\bfseries Output:} value $ v(s)$, (optional) child node
   \STATE \texttt{score} $\leftarrow 0$ 
   \STATE \texttt{depth} $ \leftarrow 0$
   \STATE $ a \leftarrow $ \texttt{""} 
   \WHILE{\texttt{score} = 0 and \texttt{depth} $<  L$}
   \STATE $ a \leftarrow a + \texttt{LLM}(s + a)$
   \STATE \texttt{score} $\leftarrow$ \texttt{Verifier} $(s + a)$
   \STATE \texttt{depth} $ \leftarrow \texttt{depth} +1$
   \ENDWHILE
   \IF{\texttt{score} $ = -1$ or \texttt{depth} $ = L$}
   \STATE {\bfseries return} $ -1, \texttt{None}$
   \ELSE
   \STATE {\bfseries return} $ +1, s+a$
   \ENDIF
   
   
\end{algorithmic}
   \caption{Evaluate and (maybe) expand}
   \label{alg:eval_and_expand}
\end{algorithm}
\end{minipage}
\end{wrapfigure}

where $p_s$ is the prior at this node, $c_\textit{UCT}$ is a global exploration coefficient, $N_\textit{parent}$ is the number of visits at the parent node, $N_s$ is the number of visits at this node, and $v_i$ is the estimated value at the $ i$th visit to $s$. Note that this selection procedure has two hyperparameters: $ c_{UCT}$ and $ p_{widen}$ that encourage selecting more rarely visited nodes and widen nodes respectively. 

\paragraph{Combining expansion and evaluation.} 
Traditionally, MCTS will first expand a node into children and evaluate it either by simulated rollouts \citep{Chaslot2008ProgressiveSF, zhang2023planning} or a learned value function \citep{Silver2016MasteringTG}. Neither of these methods is a good fit for our problem because generating rollouts requires many expensive calls to the LLM and learning a value requires large amounts of training data. Moreover, both methods give noisy signal, but in our setting we have access to the ground truth verifier. We will use the verifier 

Beyond being noiseless, the verifier has one more important property: if a partial program fails the verifier, no subsequent completion can yield success. So, we want to make sure that we never add to the tree any expansion that is a guaranteed failure. Doing this require explicitly linking expansion to evaluation where we evaluate the node and \emph{maybe} expand it, as formalized in Algorithm \ref{alg:eval_and_expand}.

In addition to only adding nodes with potential to the tree, we want to leverage the verifier to cheaply evaluate partial programs without extra calls to the LLM. Explicitly, from a node containing the string $ s $ we continue to extend $ a$ with the LLM until the verifier is able to return a valid score. At this point, we can return the estimated value $ v(s)$ of $ s$ as follows:
\begin{align}
    v(s) = \texttt{Verifier}(s + a) = \begin{cases}
        + 1 & \text{verified, but may be incomplete.}\\
        -1 & \text{verified as a failure.}
    \end{cases}
\end{align}
If $ v(s) = +1$, we also add $ s+a$ as a child in the tree, while if $ v(s) = -1$, we do not add $ s+a$ since it is a verified failure.
Appendix~\ref{sec:scoring-exs} gives explicit examples of scoring partial programs.

\paragraph{Backpropagation.} The last step of an iteration of MCTS is to backpropagate the observed value from leaf back up to root. We do this in the standard way so that signal is propagated up the tree. The algorithm terminates when it finds a complete solution that verifies or when it exceeds some token limit or time limit. 

\subsection{Connecting the partial program score to the MDP}
Importantly, while the verifier gives us ground truth information about whether the program verifies so far, it does not give an unbiased estimate of the true value of a state in the MDP defined above. Instead, we can view our use of the verifier as a heuristic that quickly returns an \emph{upper bound} on the value function of a potential child. Recall that the value function $ V^* $ of the optimal policy in a deterministic MDP with state-based rewards like ours is defined by the Bellman equation $ V^*(s) = \max_a  r(s) + V^*(s+a)$. With this definition, we can formally describe the optimism property of our estimates values as follows:
\begin{lemma}
    The value $ v(s)$ returned by Algorithm \ref{alg:eval_and_expand} satisfies the following:
    \begin{align}
    v(s) \geq \mathbb{E}_{a \sim \texttt{LLM} + \texttt{Verifier} | s}[V^*(s+a)]
\end{align}
\end{lemma}

This is fairly straightforward to prove. If $ v(s) = -1$, then we know that the sampled completion $ a $ is a failure no matter what happens afterwards, so $ v(s) = V^*(s+a) = -1$. On the other hand, if $ v(s) = 1$ then we are assigning the maximal possible value in this MDP, so $ v(s) \geq V^*(s+a)$.  

In this way, our value estimate is explicitly an \emph{optimistic} estimate of the value. This is even beyond the UCT score computed by MCTS. We hypothesize that this encourages deeper exploration of the search trees which can be beneficial in the multi-step problems we consider.

\section{A problem suite for multi-step verified programming}

\subsection{Defining the problems}
We are not aware of any existing collections of problems that are designed for multi-step program synthesis and checked using verifiers. That is why we have created our own problem suite of nine problems. The problems represent a meaningful scenarios in verified programming. They require creating Algebraic Data Types (ADTs), defining functions on them using pattern matching, and proving properties using induction. 
Compared to prior benchmarks, the problems require more intricate multi-step reasoning and test capabilities that are specifically important for verified programming.
The problems are defined as follows:
\begin{description}
\item[Factorial] asks to define the factorial function and to prove that it is always strictly positive.
\item[Opt0] asks to define an ADT for arithmetic expressions, an optimizer, and to prove that the optimizer preserves semantics.
\item[Opt0 Opt] asks to define an ADT for arithmetic expressions, an optimizer, an optimal predicate, and to prove that the optimizer is optimal.
\item[BST] asks to define a tree, the binary search tree (BST) property, insertion, and to prove two properties of insertions (membership and BST preservation).
\item[Repeat] asks to define a function returning a list with a given element repeated a given number of times, and to prove two properties related to length and membership.
\item[Lights] asks to define an ADT for traffic lights, then write a function ensuring that red and green lights are always separated by yellow lights, and then to prove its correctness.
\item[Food] asks to define an ADT that represents different foods with toppings, and a predicate about the amount of toppings, and to prove a property of this predicate.
\item[Days] asks to define an ADT that represents days of the week, two functions that iterate through business days, and then to prove a property of weekdays.
\item[Reverse] asks to define a function that reverses a list, and prove two properties of list reversals (permutation and involution).
\end{description}
All problems are implemented in Dafny, and all but the last three are implemented in Coq, giving a total of 15 problems. Since the Coq verifier has substantially less automation than Dafny which leads to longer proofs and since the model is not always very consistent at Coq syntax, just for Coq we provide some syntax hints in the prompt. The full prompts can be found in \Cref{sec:prompts}.

\subsection{Criteria for Success}
In order to be considered successful, a program must first pass the verifier and some syntactic checks (e.g. the presence of a proof marker and a problem-specific minimum number of lines of code). These initial checks are meant to ensure the model has made a successful attempt prove a lemma.

A second check ensures that the model has proven the correct lemma: In order to check whether a model has proven a property, we inject a second lemma with it, and prove it by referring to the lemma we asked the model to write. If the model has proven this lemma as directed, this new code including check lemma will verify successfully. If the model has proven an incorrect lemma, a verifier error will be produced. 
Note that the check lemma is only injected into the verifier input. The model does not get to see it, so this check does not provide additional hints to the model.

A full description of each problem including the prompts and lemmas used for checking success can be found in \Cref{sec:prompts}.

\section{Experimental setup}

\subsection{Pass@$T$ evaluation metric}\label{sec:T}

We report all of our results in terms of pass@$T$, which is, to our knowledge, a novel metric inspired by pass@$k$ that is often used in code generation benchmarks \citep{chen2021evaluating}. While pass@$k$ computes the probability of generating a success when we sample $ k $ programs, pass@$T$ computes the probability of success if we allow the model to sample $ T $ tokens.
Pass@$T$ has several benefits:
\begin{enumerate}
    \item Pass@$T$ fairly compares methods. One run of MCTS can be much more expensive than sampling one program from a model, so using pass@$k$ is not fair. In contrast pass@$T$ really estimates the dominant cost of generation, namely how many tokens need to be generated to yield success.
    \item Pass@$T$ controls for hardware and implementation variability. Compared to using wall-clock time, using pass@$T$ does not depend on the underlying hardware and system-level optimizations. 
\end{enumerate}

To estimate pass@$T$, we generate $ n $ runs per problem of up to $ T_{max}$ tokens per run (where if the run terminates successfully before $ T_{max}$ we stop the run). Then for each $ T \leq T_{max}$, we have $ n$ binary trials indicating whether that run terminated successfully in $ \leq T$ tokens. In the results, we report the mean of these $ n $ binary variables and also 95\% Wilson intervals \citep{Wilson1927ProbableIT}.

\subsection{Base model}\label{sec:base-model}

VerMCTS is compatible with any base model and only requires sampling from the model (no training is needed). We opt to use an open-weights model as the base language model and then compare different sampling procedures on top of this base model. Specifically, we use Phind-CodeLLama-34B-v2 \citep{phind, roziere2023code}. This models has been trained explicitly for code generation, but the verified programming languages we use are relatively ``low resource'' languages, so the models will perform worse than at high-resource languages \citep{cassano:multipl-e}.

\subsection{Baselines}

We consider a variety of baseline methods to illustrate the benefits of leveraging the verifier inside of VerMCTS. 
\begin{itemize}
    \item \textbf{Whole sampling.} The most naive baseline just samples entire programs from the base model. To compute pass@$T$ we just continue generating new samples until success or until the token limit is reached.
    \item \textbf{Rollout MCTS.} Related work on MCTS uses rollouts to evaluate a node \citep{Chaslot2008ProgressiveSF, zhang2023planning}. We ablate the importance of using the verifier by replacing the ``evaluate and maybe expand'' step with separate expand and evaluate steps. We expand by sampling a fixed number of actions $ k $ from the LLM and evaluate by rolling out with the LLM to a terminal node before querying the reward function.
    \item \textbf{Reflexion.} Finally, to show how VerMCTS is efficient at incorporating information from the verifier we also compare to a Reflexion \citep{shinn2023reflexion} baseline where the LLM gets to view the errors produced by the verifier on failed attempts. 
\end{itemize}

\subsection{Hyperparamters}
There are a few important hyperparameters. Shared across all methods are the LLMs which require some sampling hyperparameters. For all methods we use nucleus sampling \citet{holtzman2019curious} with top-p=0.95 following \citet{roziere2023code}. For every method, we sweep over temperature on one representative problem and use that temperature for the rest. Our VerMCTS algorithm also introduces two hyperparameters that govern exploration: $ c_{UCT}$ and $ p_{widen}$ which we found fairly straightforward to set. Full discussion of the hyperparameters can be found in \Cref{sec:setup}.

\section{Results}

\begin{figure}[t]
    \centering
    \subfigure[Dafny]{%
        \includegraphics[width=0.45\textwidth]{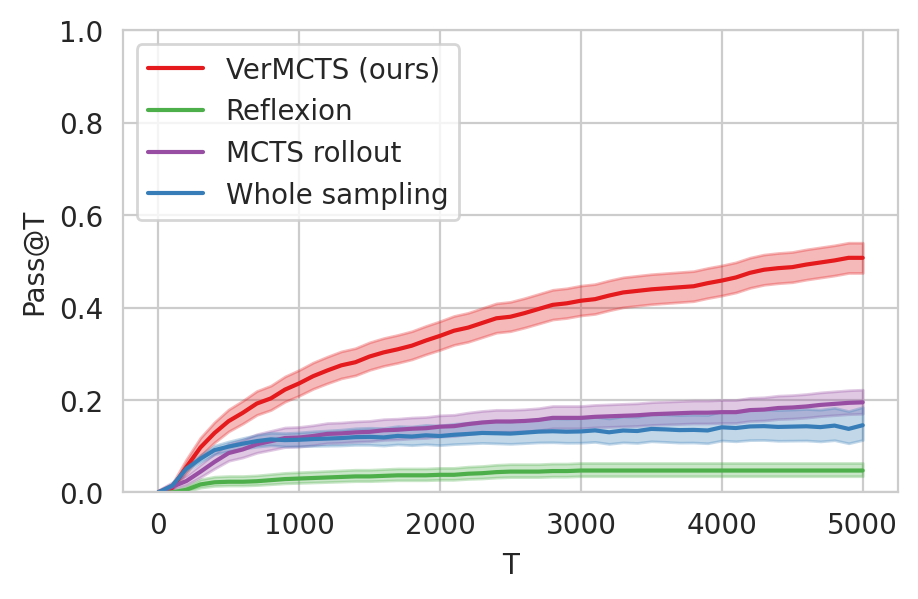}
        \label{fig:dafny_avg}
    }
    \hspace{0.1cm} 
    \subfigure[Coq]{%
        \includegraphics[width=0.45\textwidth]{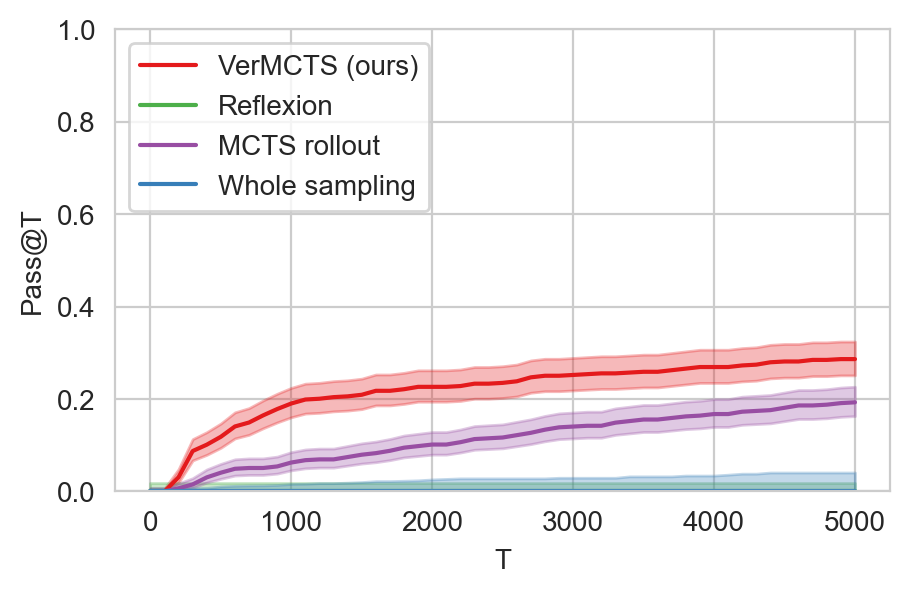}
        \label{fig:coq_avg}
    }
    \vspace{-0.2cm}
    \caption{Average results for pass@T vs. T (the number of tokens) for various baseline methods on our suite of programming problems in Dafny and Coq.}
    \label{fig:avg}
\end{figure}


\subsection{Synthesizing programs in Dafny and Coq}

We run VerMCTS and our three baselines across our full suite of problems. The aggregate results are illustrated in \Cref{fig:avg}. In both programming languages VerMCTS convincingly outperforms the baselines. Generally, MCTS rollout is second best, followed by whole sampling and then Reflexion. As previewed in the introduction, we see about a 30\% absolute improvement in pass@5000 for VerMCTS relative to whole sampling. This amounts to a 4x relative improvement on Dafny and on Coq, whole sampling does not even get a pass rate above 0 at this budget. Note that Coq is substantially more challenging since the verifier is less automated, making the proofs more involved.

Examining the performance of the baselines more closely, we see that MCTS rollout does outperform whole sampling, suggesting that there are some gains to be had from not re-generating from scratch every time even if the verifier is not used to guide the search at intermediate steps. But, using the verifier in VerMCTS provides even better performance. Looking at Reflexion, we see that performance is poor on these tasks. This could be due to many reasons including: (1) the base model is not good at responding to errors in low resource languages like Dafny and Coq, (2) the base model does not do well integrating the long contexts created by the Reflexion prompts, and (3) Reflexion does not make it as easy to backtrack once the model gets stuck with a particular error.

\begin{figure}[b]
    \centering
    \includegraphics[width=\textwidth]{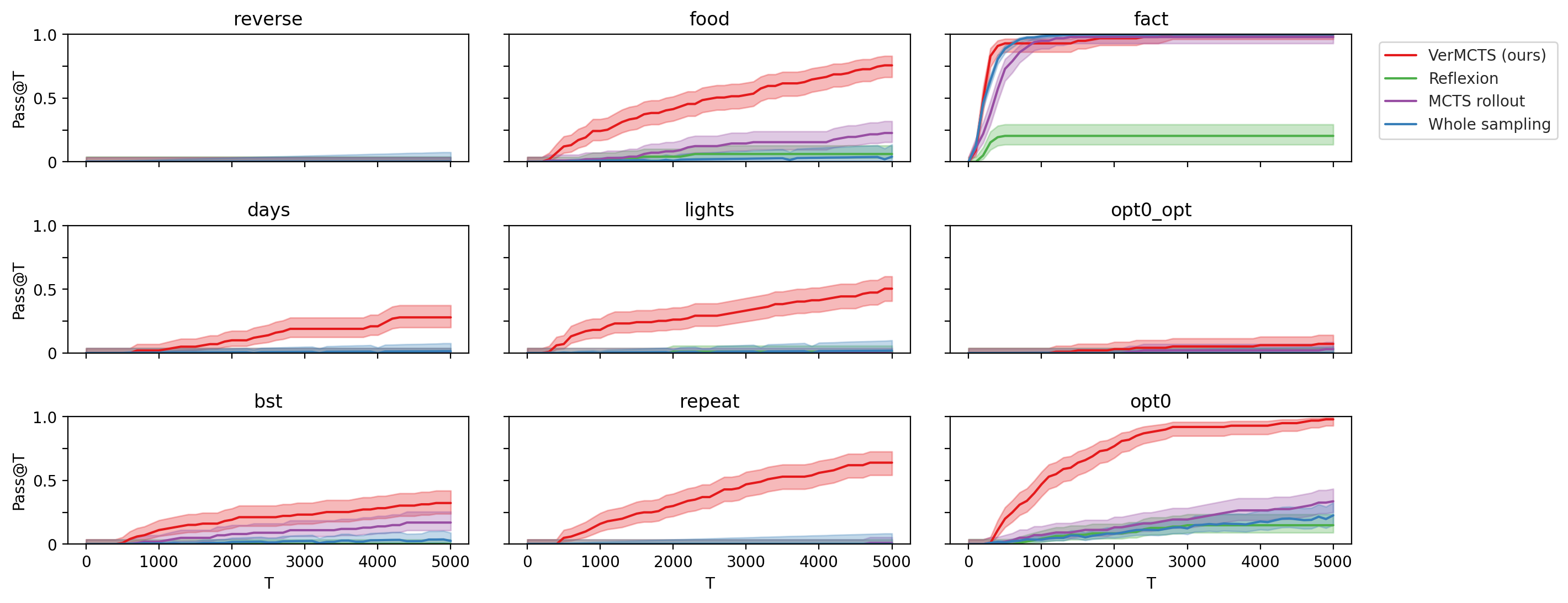}
    \vspace{-0.6cm}
    \caption{Pass@T results for all algorithms on our suite of problems in Dafny.}
    \label{fig:dafny_all}
\end{figure}

\begin{figure}[t]
    \centering
    \includegraphics[width=\textwidth]{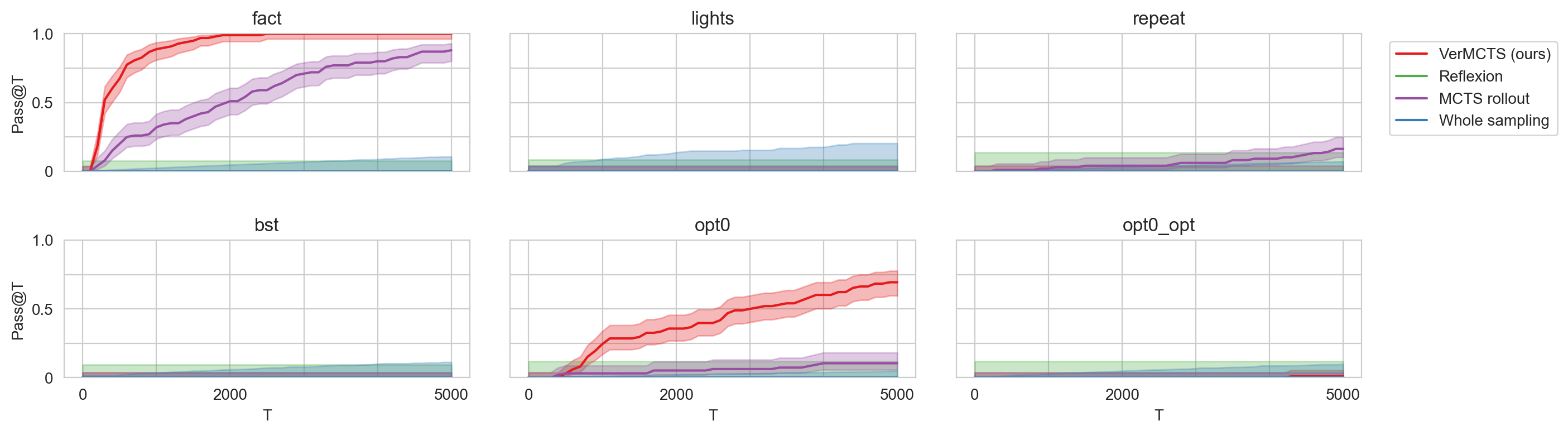}
    \vspace{-0.6cm}
    \caption{Pass@T results for all algorithms on our suite of problems in Coq.}
    \label{fig:coq_all}
\end{figure}

In Figures \Cref{fig:dafny_all} and \Cref{fig:coq_all}, we present the per-problem results on our problem suite. There is substantial variation across problems, but across all problems VerMCTS is the best approach or within the margin of error, often exceeding the baselines by a large margin and sometimes solving problems that no baseline solves at all. That said, some problems are clearly challenging: on one problem in Dafny and three in Coq, none of the algorithms find a solution within 5000 tokens.

\subsection{Examining the VerMCTS search trees}

\begin{figure}[b]
    \centering
    \includegraphics[width=\textwidth]{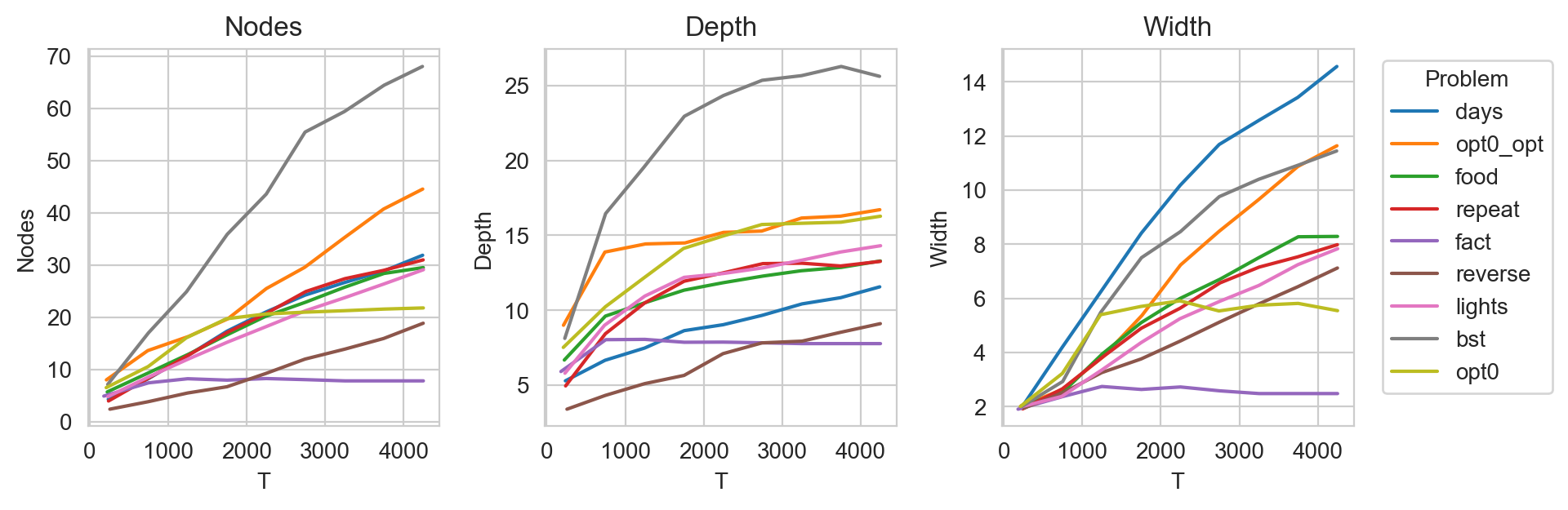}
    \vspace{-0.6cm}
    \caption{Average number of nodes, depth, and width of the VerMCTS search tree as the number of tokens increases across the full suite of Dafny problems. Recall that failed expansions are not added to the tree. Harder problems tend to lead to larger trees.}
    \label{fig:trees}
\end{figure}

In \Cref{fig:trees} we provide an experiment to probe for a mechanistic understanding of how VerMCTS works in Dafny. We consider the number of nodes (excluding widen nodes), the depth and the width of the search trees as the number of tokens generated increases. Note that since we do not add failed expansions to the tree, sometimes more tokens are generated without adding nodes to the tree. Generally, we observe that the more challenging problems (with lower pass rates) tend to lead to larger search trees, indicating that the algorithm is successfully. We also notice that while the number of nodes grows fairly linearly across time for most problems, the depth grows earlier and then flattens out. This suggests that the VerMCTS search is closer to ``depth first``, first pushing an expansion branch to a terminal node before going back and widening the tree.

\section{Related Work}

\paragraph{Neural Program Synthesis with Large Language Models}
\citet{austin2021program} and \citet{chen2021evaluating} demonstrated that Large Language Models (LLMs) can generate correct Python programs from natural language descriptions. These studies introduced the MBPP and HumanEval datasets, respectively, which are widely used for evaluating LLMs in program synthesis tasks.
\citet{cassano:multipl-e} extended this concept by showing that LLMs can also generate programs in over 20 languages other than Python. This was achieved by translating the MBPP and HumanEval datasets using their system, MultiPL-E. Their findings indicate that generating accurate programs in lower resource languages is more challenging compared to higher resource languages, such as Python. 
In our experiments, for proof synthesis, we have another dimension of challenge: some languages (Coq) are inherently more challenging than others (Dafny), depending on how much automation the verifiers provide.
However, none of these works explored the generation of programs that are correct by construction.

\paragraph{Symbolic Algorithms for Neural Program Synthesis}
\citet{grand2023lilo} integrated a classic symbolic top-down synthesis algorithm for library learning~\cite{topdown-synthesis} with LLMs. \citet{opentau} employed program decomposition and a bottom-up tree-search algorithm to infer missing TypeScript types. 
\citet{zhou2023language} used Monte Carlo Tree Search (MCTS) to create single-function programs in Python. 
\citet{zhang2023planning} applied a tree-based planning algorithm for decoding LLM token sequences, which were then evaluated for correctness using a test suite.
\citet{lample2022hypertree} adapted MCTS for neural theorem proving by employing a tree-based search algorithm to generate proof trees in Lean.
Different from these closely related works, we (1) focus on verified program synthesis in Dafny and Coq, and (2) leverage the verifier inside the loop of the search algorithm to efficiently guide the search.

\paragraph{Theorem Proving with Large Language Models}
\citet{han2022proof} demonstrated that LLMs can be trained to generate proofs in Lean through self-supervision. \citet{yang2023leandojo} presented that Retrieval-Augmented Generation (RAG)~\cite{re2g} models significantly enhance LLMs' performance in theorem proving tasks. 
\citet{baldur} employed a methodology akin to that of \citet{han2022proof} to generate and repair complete proofs in Isabelle/HOL. \citet{jiang2023draft} introduced methods to first map natural language proofs to formal proof sketches in Isabelle and then fill in the gaps using an automated prover.
These studies predominantly used LLMs to iteratively generate individual proof steps, which were then verified using a theorem prover.
\citet{thakur2023languageagent} propose a language-agent approach to formal theorem-proving, alternating selection and execution steps.
In contrast, we focus on verified program synthesis and developing a method that effectively integrates a verifier and LLM without any additional training.

\paragraph{Scoring Partial Programs}
\citet{desai2015program}, one of the first to 
effectively tackle the problem of program synthesis using natural language, used a scoring function to rank candidate partial programs. \citet{opentau} similarly used a scoring function to rank candidate
partial programs based on their types
in order to aid the tree search process, and provided multiple solutions to the user ranked by their score.
\citet{ye2021} used abstract interpretation to rule out partial programs that do not satisfy some constraints, typically on input/output examples.
\citet{chen2022codet} used LLM-generated unit tests suites and their pass rates to score candidate programs,
and provided the user with the top-scoring program.
\citet{ni2023lever} further utilized execution information to rank candidate programs.
\citet{shirafuji2023refactoring} used a scoring function to rank example refactoring programs generated by an LLM
before applying them to the given code.
\citet{coder-reviewer} studies using scoring functions to rank candidate partial programs in-depth,
and proposes the use of a \textit{reviewer} model to score candidate programs based on how closely they
match the given instruction.
Most of these works have scored partial programs specified as grammatical programs with holes as opposed to our left-to-right generation of partial programs, and have not considered verified programming languages.

\section{Discussion}

We have demonstrated that relatively weak language models can reliably produce verified code by guiding a search process that verifies partial programs at each step.
Our technique shines on multi-step problems, made of dependent sub-problems.
Our technique can be adapted to a setting where the interfaces and specifications are given, and the code is verified at each step by additional code containing assertions or proofs.

\paragraph{Limitations.} 
A key aspect of our approach resides in the scoring of partial programs.
However, the scoring is limited by coarse granularity and lack of lookahead in the scoring function.
The granularity of the verification step is a whole unit, e.g. a function in Dafny and a command in Coq. For Dafny, the coarse granularity means we have to wait multiple lines to get feedback. For Coq, the fine granularity doesn't help much with bigger proofs, which require planning.

\paragraph{Future work.}
What we find most interesting and promising about our approach is that so much is possible by a ``blind'' search that only uses scalar reward signal.
In future work, it would be fruitful to find ways of allowing the search to rely on richer feedback while maintaining the efficiency of leveraging the verifier to avoid doing costly rollouts or reflection steps. 
Moreover, it will be interesting to see if the basic idea of VerMCTS, using a cheap and provable upper bound on the value function to guide search, can be applied beyond the verified programming setting.

\unless\ifanon
\section*{Acknowledgments}
We thank Matko Bošnjak, Kevin Ellis, Samy Jelassi, Gabriel Poesia, and Garrett Tanzer for insightful discussions, and Vivan Hui and Lisa Zhang for suggestions on drafts.
We thank the Harvard SEAS Computing group for help in access to GPUs.
Tarun Prasad, Chloe Loughridge, and Sabrina Ruixin Hu were partially supported by the Harvard College Research Program (HCRP) through the Harvard College Office of Undergraduate Research and Fellowships.
Nada Amin was partially supported by NSF Award 2303983.
Support for William E. Byrd's work on this paper was provided by NCATS, through the Biomedical Data Translator program (NIH award OT2TR003435).
Any opinions expressed in this document do not necessarily reflect the views 
of NSF, NCATS, individual Translator team members, or affiliated 
organizations and institutions.
\fi

\bibliographystyle{plainnat}
\bibliography{paper}

\begin{thebibliography}{33}
\providecommand{\natexlab}[1]{#1}
\providecommand{\url}[1]{\texttt{#1}}
\expandafter\ifx\csname urlstyle\endcsname\relax
  \providecommand{\doi}[1]{doi: #1}\else
  \providecommand{\doi}{doi: \begingroup \urlstyle{rm}\Url}\fi

\bibitem[Austin et~al.(2021)Austin, Odena, Nye, Bosma, Michalewski, Dohan, Jiang, Cai, Terry, Le, and Sutton]{austin2021program}
Jacob Austin, Augustus Odena, Maxwell Nye, Maarten Bosma, Henryk Michalewski, David Dohan, Ellen Jiang, Carrie Cai, Michael Terry, Quoc Le, and Charles Sutton.
\newblock Program synthesis with large language models.
\newblock \emph{arXiv preprint arXiv:2108.07732}, 2021.

\bibitem[Bowers et~al.(2023)Bowers, Olausson, Wong, Grand, Tenenbaum, Ellis, and Solar-Lezama]{topdown-synthesis}
Matthew Bowers, Theo~X. Olausson, Lionel Wong, Gabriel Grand, Joshua~B. Tenenbaum, Kevin Ellis, and Armando Solar-Lezama.
\newblock Top-down synthesis for library learning.
\newblock \emph{Proc. ACM Program. Lang.}, 7\penalty0 (POPL), jan 2023.
\newblock \doi{10.1145/3571234}.
\newblock URL \url{https://doi.org/10.1145/3571234}.

\bibitem[Cassano et~al.(2023{\natexlab{a}})Cassano, Gouwar, Nguyen, Nguyen, {Phipps-Costin}, Pinckney, Yee, Zi, Anderson, Feldman, Guha, Greenberg, and Jangda]{cassano:multipl-e}
Federico Cassano, John Gouwar, Daniel Nguyen, Sydney Nguyen, Luna {Phipps-Costin}, Donald Pinckney, Ming-Ho Yee, Yangtian Zi, Carolyn~Jane Anderson, Molly~Q. Feldman, Arjun Guha, Michael Greenberg, and Abhinav Jangda.
\newblock {{MultiPL-E}}: {{A Scalable}} and {{Polyglot Approach}} to {{Benchmarking Neural Code Generation}}.
\newblock \emph{IEEE Transactions on Software Engineering (TSE)}, 49\penalty0 (7):\penalty0 3675--3691, 2023{\natexlab{a}}.

\bibitem[Cassano et~al.(2023{\natexlab{b}})Cassano, Yee, Shinn, Guha, and Holtzen]{opentau}
Federico Cassano, Ming-Ho Yee, Noah Shinn, Arjun Guha, and Steven Holtzen.
\newblock Type prediction with program decomposition and fill-in-the-type training, 2023{\natexlab{b}}.

\bibitem[Chaslot et~al.(2008)Chaslot, Winands, Herik, Uiterwijk, and Bouzy]{Chaslot2008ProgressiveSF}
Guillaume Chaslot, Mark H.~M. Winands, H~Jaap Van~Den Herik, Jos Uiterwijk, and Bruno Bouzy.
\newblock Progressive strategies for monte-carlo tree search.
\newblock \emph{New Mathematics and Natural Computation}, 04:\penalty0 343--357, 2008.
\newblock URL \url{https://api.semanticscholar.org/CorpusID:1719063}.

\bibitem[Chen et~al.(2022)Chen, Zhang, Nguyen, Zan, Lin, Lou, and Chen]{chen2022codet}
Bei Chen, Fengji Zhang, Anh Nguyen, Daoguang Zan, Zeqi Lin, Jian-Guang Lou, and Weizhu Chen.
\newblock Codet: Code generation with generated tests, 2022.

\bibitem[Chen et~al.(2021)Chen, Tworek, Jun, Yuan, Pinto, Kaplan, Edwards, Burda, Joseph, Brockman, et~al.]{chen2021evaluating}
Mark Chen, Jerry Tworek, Heewoo Jun, Qiming Yuan, Henrique Ponde de~Oliveira Pinto, Jared Kaplan, Harri Edwards, Yuri Burda, Nicholas Joseph, Greg Brockman, et~al.
\newblock Evaluating large language models trained on code.
\newblock \emph{arXiv preprint arXiv:2107.03374}, 2021.

\bibitem[Cou{\"e}toux et~al.(2011)Cou{\"e}toux, Hoock, Sokolovska, Teytaud, and Bonnard]{Coutoux2011ContinuousUC}
Adrien Cou{\"e}toux, Jean-Baptiste Hoock, Nataliya Sokolovska, Olivier Teytaud, and Nicolas Bonnard.
\newblock Continuous upper confidence trees.
\newblock In \emph{Learning and Intelligent Optimization}, 2011.
\newblock URL \url{https://api.semanticscholar.org/CorpusID:13463524}.

\bibitem[Desai et~al.(2016)Desai, Gulwani, Hingorani, Jain, Karkare, Marron, R, and Roy]{desai2015program}
Aditya Desai, Sumit Gulwani, Vineet Hingorani, Nidhi Jain, Amey Karkare, Mark Marron, Sailesh R, and Subhajit Roy.
\newblock Program synthesis using natural language.
\newblock In \emph{Proceedings of the 38th International Conference on Software Engineering}, ICSE '16, page 345–356, New York, NY, USA, 2016. Association for Computing Machinery.
\newblock ISBN 9781450339001.
\newblock \doi{10.1145/2884781.2884786}.
\newblock URL \url{https://doi.org/10.1145/2884781.2884786}.

\bibitem[First et~al.(2023)First, Rabe, Ringer, and Brun]{baldur}
Emily First, Markus Rabe, Talia Ringer, and Yuriy Brun.
\newblock Baldur: Whole-proof generation and repair with large language models.
\newblock In \emph{Proceedings of the 31st ACM Joint European Software Engineering Conference and Symposium on the Foundations of Software Engineering}, ESEC/FSE 2023, page 1229–1241, New York, NY, USA, 2023. Association for Computing Machinery.
\newblock ISBN 9798400703270.
\newblock \doi{10.1145/3611643.3616243}.
\newblock URL \url{https://doi.org/10.1145/3611643.3616243}.

\bibitem[Glass et~al.(2022)Glass, Rossiello, Chowdhury, Naik, Cai, and Gliozzo]{re2g}
Michael Glass, Gaetano Rossiello, Md~Faisal~Mahbub Chowdhury, Ankita Naik, Pengshan Cai, and Alfio Gliozzo.
\newblock {R}e2{G}: Retrieve, rerank, generate.
\newblock In Marine Carpuat, Marie-Catherine de~Marneffe, and Ivan~Vladimir Meza~Ruiz, editors, \emph{Proceedings of the 2022 Conference of the North American Chapter of the Association for Computational Linguistics: Human Language Technologies}, pages 2701--2715, Seattle, United States, July 2022. Association for Computational Linguistics.
\newblock \doi{10.18653/v1/2022.naacl-main.194}.
\newblock URL \url{https://aclanthology.org/2022.naacl-main.194}.

\bibitem[Grand et~al.(2023)Grand, Wong, Bowers, Olausson, Liu, Tenenbaum, and Andreas]{grand2023lilo}
Gabriel Grand, Lionel Wong, Matthew Bowers, Theo~X. Olausson, Muxin Liu, Joshua~B. Tenenbaum, and Jacob Andreas.
\newblock Lilo: Learning interpretable libraries by compressing and documenting code, 2023.

\bibitem[Han et~al.(2022)Han, Rute, Wu, Ayers, and Polu]{han2022proof}
Jesse~Michael Han, Jason Rute, Yuhuai Wu, Edward Ayers, and Stanislas Polu.
\newblock Proof artifact co-training for theorem proving with language models.
\newblock In \emph{International Conference on Learning Representations}, 2022.
\newblock URL \url{https://openreview.net/forum?id=rpxJc9j04U}.

\bibitem[Holtzman et~al.(2019)Holtzman, Buys, Du, Forbes, and Choi]{holtzman2019curious}
Ari Holtzman, Jan Buys, Li~Du, Maxwell Forbes, and Yejin Choi.
\newblock The curious case of neural text degeneration.
\newblock \emph{arXiv preprint arXiv:1904.09751}, 2019.

\bibitem[ImparaAI(2024)]{impara-mcts}
ImparaAI.
\newblock Monte carlo tree search.
\newblock \url{https://github.com/ImparaAI/monte-carlo-tree-search}, 2024.

\bibitem[Jiang et~al.(2023)Jiang, Welleck, Zhou, Lacroix, Liu, Li, Jamnik, Lample, and Wu]{jiang2023draft}
Albert~Qiaochu Jiang, Sean Welleck, Jin~Peng Zhou, Timothee Lacroix, Jiacheng Liu, Wenda Li, Mateja Jamnik, Guillaume Lample, and Yuhuai Wu.
\newblock Draft, sketch, and prove: Guiding formal theorem provers with informal proofs.
\newblock In \emph{The Eleventh International Conference on Learning Representations}, 2023.
\newblock URL \url{https://openreview.net/forum?id=SMa9EAovKMC}.

\bibitem[Lample et~al.(2022)Lample, Lacroix, anne Lachaux, Rodriguez, Hayat, Lavril, Ebner, and Martinet]{lample2022hypertree}
Guillaume Lample, Timothee Lacroix, Marie anne Lachaux, Aurelien Rodriguez, Amaury Hayat, Thibaut Lavril, Gabriel Ebner, and Xavier Martinet.
\newblock Hypertree proof search for neural theorem proving.
\newblock In Alice~H. Oh, Alekh Agarwal, Danielle Belgrave, and Kyunghyun Cho, editors, \emph{Advances in Neural Information Processing Systems}, 2022.
\newblock URL \url{https://openreview.net/forum?id=J4pX8Q8cxHH}.

\bibitem[Ni et~al.(2023)Ni, Iyer, Radev, Stoyanov, Yih, Wang, and Lin]{ni2023lever}
Ansong Ni, Srini Iyer, Dragomir Radev, Ves Stoyanov, Wen-tau Yih, Sida~I Wang, and Xi~Victoria Lin.
\newblock Lever: Learning to verify language-to-code generation with execution.
\newblock In \emph{Proceedings of the 40th International Conference on Machine Learning (ICML'23)}, 2023.

\bibitem[Phind(2023)]{phind}
Phind.
\newblock Beating gpt-4 on humaneval with a fine-tuned codellama-34b.
\newblock \url{https://www.phind.com/blog/code-llama-beats-gpt4}, 2023.

\bibitem[Rosin(2011)]{Rosin2011MultiarmedBW}
Christopher~D. Rosin.
\newblock Multi-armed bandits with episode context.
\newblock \emph{Annals of Mathematics and Artificial Intelligence}, 61:\penalty0 203--230, 2011.
\newblock URL \url{https://api.semanticscholar.org/CorpusID:207081359}.

\bibitem[Roziere et~al.(2023)Roziere, Gehring, Gloeckle, Sootla, Gat, Tan, Adi, Liu, Remez, Rapin, et~al.]{roziere2023code}
Baptiste Roziere, Jonas Gehring, Fabian Gloeckle, Sten Sootla, Itai Gat, Xiaoqing~Ellen Tan, Yossi Adi, Jingyu Liu, Tal Remez, J{\'e}r{\'e}my Rapin, et~al.
\newblock Code llama: Open foundation models for code.
\newblock \emph{arXiv preprint arXiv:2308.12950}, 2023.

\bibitem[Shinn et~al.(2023)Shinn, Cassano, Gopinath, Narasimhan, and Yao]{shinn2023reflexion}
Noah Shinn, Federico Cassano, Ashwin Gopinath, Karthik~R Narasimhan, and Shunyu Yao.
\newblock Reflexion: Language agents with verbal reinforcement learning.
\newblock In \emph{Thirty-seventh Conference on Neural Information Processing Systems}, 2023.

\bibitem[Shirafuji et~al.(2023)Shirafuji, Oda, Suzuki, Morishita, and Watanobe]{shirafuji2023refactoring}
Atsushi Shirafuji, Yusuke Oda, Jun Suzuki, Makoto Morishita, and Yutaka Watanobe.
\newblock Refactoring programs using large language models with few-shot examples, 2023.

\bibitem[Silver et~al.(2016)Silver, Huang, Maddison, Guez, Sifre, van~den Driessche, Schrittwieser, Antonoglou, Panneershelvam, Lanctot, Dieleman, Grewe, Nham, Kalchbrenner, Sutskever, Lillicrap, Leach, Kavukcuoglu, Graepel, and Hassabis]{Silver2016MasteringTG}
David Silver, Aja Huang, Chris~J. Maddison, Arthur Guez, L.~Sifre, George van~den Driessche, Julian Schrittwieser, Ioannis Antonoglou, Vedavyas Panneershelvam, Marc Lanctot, Sander Dieleman, Dominik Grewe, John Nham, Nal Kalchbrenner, Ilya Sutskever, Timothy~P. Lillicrap, Madeleine Leach, Koray Kavukcuoglu, Thore Graepel, and Demis Hassabis.
\newblock Mastering the game of go with deep neural networks and tree search.
\newblock \emph{Nature}, 529:\penalty0 484--489, 2016.
\newblock URL \url{https://api.semanticscholar.org/CorpusID:515925}.

\bibitem[Thakur et~al.(2023)Thakur, Wen, and Chaudhuri]{thakur2023languageagent}
Amitayush Thakur, Yeming Wen, and Swarat Chaudhuri.
\newblock A language-agent approach to formal theorem-proving, 2023.

\bibitem[Wilson(1927)]{Wilson1927ProbableIT}
Edwin~Bidwell Wilson.
\newblock Probable inference, the law of succession, and statistical inference.
\newblock \emph{Journal of the American Statistical Association}, 22:\penalty0 209--212, 1927.
\newblock URL \url{https://api.semanticscholar.org/CorpusID:121572396}.

\bibitem[Wolf et~al.(2020)Wolf, Debut, Sanh, Chaumond, Delangue, Moi, Cistac, Rault, Louf, Funtowicz, Davison, Shleifer, von Platen, Ma, Jernite, Plu, Xu, Scao, Gugger, Drame, Lhoest, and Rush]{wolf-etal-2020-transformers}
Thomas Wolf, Lysandre Debut, Victor Sanh, Julien Chaumond, Clement Delangue, Anthony Moi, Pierric Cistac, Tim Rault, Rémi Louf, Morgan Funtowicz, Joe Davison, Sam Shleifer, Patrick von Platen, Clara Ma, Yacine Jernite, Julien Plu, Canwen Xu, Teven~Le Scao, Sylvain Gugger, Mariama Drame, Quentin Lhoest, and Alexander~M. Rush.
\newblock Transformers: State-of-the-art natural language processing.
\newblock In \emph{Proceedings of the 2020 Conference on Empirical Methods in Natural Language Processing: System Demonstrations}, pages 38--45, Online, October 2020. Association for Computational Linguistics.
\newblock URL \url{https://www.aclweb.org/anthology/2020.emnlp-demos.6}.

\bibitem[Yang et~al.(2023)Yang, Swope, Gu, Chalamala, Song, Yu, Godil, Prenger, and Anandkumar]{yang2023leandojo}
Kaiyu Yang, Aidan Swope, Alex Gu, Rahul Chalamala, Peiyang Song, Shixing Yu, Saad Godil, Ryan Prenger, and Anima Anandkumar.
\newblock {LeanDojo}: Theorem proving with retrieval-augmented language models.
\newblock In \emph{Neural Information Processing Systems (NeurIPS)}, 2023.

\bibitem[Ye et~al.(2021)Ye, Chen, Dillig, and Durrett]{ye2021}
Xi~Ye, Qiaochu Chen, Isil Dillig, and Greg Durrett.
\newblock Optimal neural program synthesis from multimodal specifications.
\newblock In Marie-Francine Moens, Xuanjing Huang, Lucia Specia, and Scott Wen-tau Yih, editors, \emph{Findings of the Association for Computational Linguistics: EMNLP 2021}, pages 1691--1704, Punta Cana, Dominican Republic, November 2021. Association for Computational Linguistics.
\newblock \doi{10.18653/v1/2021.findings-emnlp.146}.
\newblock URL \url{https://aclanthology.org/2021.findings-emnlp.146}.

\bibitem[Zhang et~al.(2023{\natexlab{a}})Zhang, Chen, Shen, Ding, Tenenbaum, and Gan]{zhang2023planning}
Shun Zhang, Zhenfang Chen, Yikang Shen, Mingyu Ding, Joshua~B. Tenenbaum, and Chuang Gan.
\newblock Planning with large language models for code generation.
\newblock In \emph{The Eleventh International Conference on Learning Representations}, 2023{\natexlab{a}}.
\newblock URL \url{https://openreview.net/forum?id=Lr8cOOtYbfL}.

\bibitem[Zhang et~al.(2023{\natexlab{b}})Zhang, Yu, Hashimoto, Lewis, Yih, Fried, and Wang]{coder-reviewer}
Tianyi Zhang, Tao Yu, Tatsunori~B. Hashimoto, Mike Lewis, Wen-tau Yih, Daniel Fried, and Sida~I. Wang.
\newblock Coder reviewer reranking for code generation.
\newblock In \emph{Proceedings of the 40th International Conference on Machine Learning}, ICML'23. JMLR.org, 2023{\natexlab{b}}.

\bibitem[Zhong and Wang(2023)]{zhong2023study}
Li~Zhong and Zilong Wang.
\newblock A study on robustness and reliability of large language model code generation.
\newblock \emph{arXiv preprint arXiv:2308.10335}, 2023.

\bibitem[Zhou et~al.(2023)Zhou, Yan, Shlapentokh-Rothman, Wang, and Wang]{zhou2023language}
Andy Zhou, Kai Yan, Michal Shlapentokh-Rothman, Haohan Wang, and Yu-Xiong Wang.
\newblock Language agent tree search unifies reasoning acting and planning in language models, 2023.

\end{thebibliography}

\newpage
\appendix

\section{Experimental Setup}\label{sec:setup}

We use the Transformers library~\cite{wolf-etal-2020-transformers} to query the LLMs.
For the MCTS, we adapt a generic open-source library~\cite{impara-mcts}.

When sampling from the LLM, we always use nucleus sampling \citep{holtzman2019curious} with $ p = 0.95$ following \cite{roziere2023code}. We tune hyperparameters on one particular problem (opt0) in Dafny, but only checking for verification and not additionally checking for correctness. Each method has slightly different hyperparameters, but we generally tune temperature of the LLM, the MCTS exploration coefficient, and the MCTS prior for widen nodes. Hyperaparameters are then fixed for all other experiments. Each algorithm's parameters are described below.

\begin{figure}
    \centering
    \includegraphics[width=0.32\textwidth]{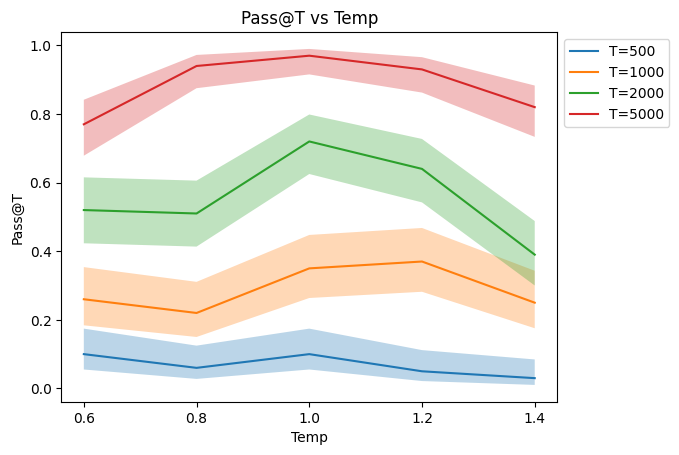}
    \includegraphics[width=0.32\textwidth]{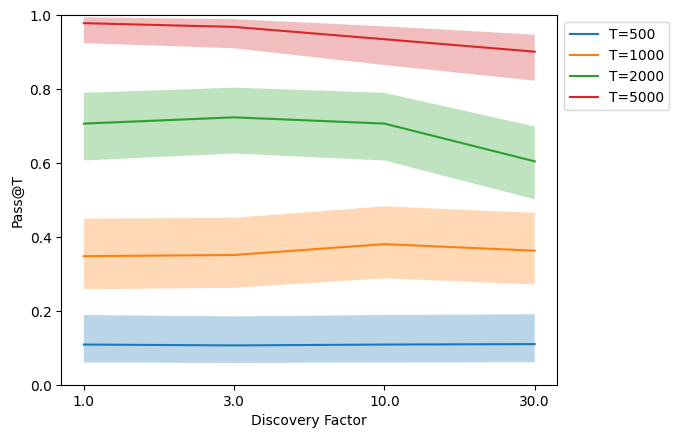}
    \includegraphics[width=0.32\textwidth]{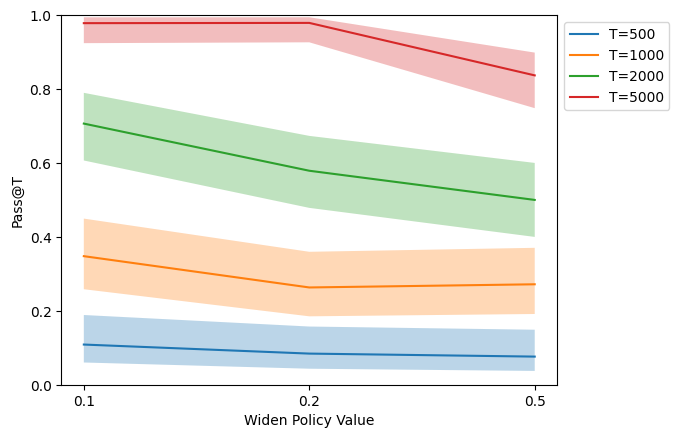}
    \caption{Hyperparameter ablations for VerMCTS on opt0 in Dafny. We find that perfromance is generally fairly stable to hyperparameter choices.}
    \label{fig:hyperparams}
\end{figure}

\paragraph{VerMCTS.} We sweep over temperature in [0.6, 0.8, 1.0, 1.0, 1.4] and find 1.0 to be best, exploration coefficient in [1, 3, 10, 30] and find 3 to be best, and the ``widen policy value'', i.e. the prior value of the widen nodes in [0.1, 0.2, 0.5] and find 0.1 to be best. See \Cref{fig:hyperparams}.

\paragraph{MCTS rollout.} We also sweep over temperature in [0.6, 0.8, 1.0, 1.0, 1.4] and find 0.8 to be best and exploration coefficient in [1, 3, 10, 30] and find 1 to be best. Note, instead of widen nodes, each node has a fixed number of children (3 in our experiments).

\paragraph{Reflexion.} We sweep over temperature in [0.2, 0.4, 0.6, 0.8, 1.0] and find 0.4 to be best.

\paragraph{Whole sampling.} We sweep over temperature in [0.2, 0.4, 0.6, 0.8, 1.0] and find 0.6 to be best.

\section{Prompts}\label{sec:prompts}
\subsection{Repeat Prompt}
\textbf{Coq.}
{\it
In Coq:
(1) Write a function `repeat` that takes an integer `x` and a natural number `n` as inputs, and returns a list of length `n` in which every element is `x`.
(2) Then write a lemma `repeat\_correct` that checks that for any `x` and `n`, `repeat` returns a list of length `n` and that every element of the list is `x`.

}
\textbf{Dafny.}
{\it
In Dafny:
(1) Write a function `repeat` that takes an integer `x` and a natural number `n` as inputs, and returns a list of length `n` in which every element is `x`.
(2) Then write a lemma `repeat\_correct` that checks that for any `x` and `n`, `repeat` returns a list of length `n` and that every element of the list is `x`.

}
\paragraph{Hints for Coq.}{\ \\
{\it
\#\#\# Hint: Start with `Require Import List. Import ListNotations.`
}}
\paragraph{Check lemma for Coq.}{\
{\begin{lstlisting}[language=dafny]
Lemma CHECK_repeat_correct: forall (x: int) (n: nat),
      length (repeat x n) = n /      forall i, 0 <= i -> i < n -> nth (repeat x n) i = x.
    Proof.
      intros.
      eapply repeat_correct; eauto.
    Qed.
\end{lstlisting}
}}
\paragraph{Check lemma for Dafny.}{\
{\begin{lstlisting}[language=dafny]
lemma CHECK_repeat_correct(x: int, n: nat)
      ensures |repeat(x, n)| == n
      ensures forall i :: 0 <= i < n ==> repeat(x, n)[i] == x
    {
      repeat_correct(x, n);
    }
\end{lstlisting}
}}
\subsection{Opt0 Opt Prompt}
\textbf{Coq.}
{\it
In Coq, write an ADT `Expr` for arithmetic expressions comprising constants, variables and binary addition. Then write a predicate `optimal` that holds on an expression if it has no additions by 0. Then write an optimizer `optimize` that removes all additions by 0. Then write a lemma `OptimizerOptimal` that ensures `optimal(optimize(e))` for all expressions `e`.

}
\textbf{Dafny.}
{\it
In Dafny, write an ADT `Expr` for arithmetic expressions comprising constants, variables and binary addition. Then write a predicate `optimal` that holds on an expression if it has no additions by 0. Then write an optimizer `optimize` that removes all additions by 0. Then write a lemma `OptimizerOptimal` that ensures `optimal(optimize(e))` for all expressions `e`.

}
\paragraph{Hints for Coq.}{\ \\
{\it
\#\#\# Hint: In the addition case, the `optimize` function should recursively optimize the sub-expressions and then match on the optimized sub-expressions.\\
\#\#\# Hint: You can import the `string` datatype with the line `Require Import Coq.Strings.String.`\\
\#\#\# Hint: Use Fixpoint instead of Definition for recursive functions.\\
\#\#\# Hint: If you do induction on `e` with sub-expressions `e1` and `e2`, the two inductive hypotheses are called `IHe1` and `IHe2`.
}}
\paragraph{Check lemma for Coq.}{\
{\begin{lstlisting}[language=dafny]
lemma CHECK_OptimizerOptimal(e: Expr) ensures optimal(optimize(e)) { OptimizerOptimal(e); }
\end{lstlisting}
}}
\paragraph{Check lemma for Dafny.}{\
{\begin{lstlisting}[language=dafny]
lemma CHECK_OptimizerOptimal(e: Expr) ensures optimal(optimize(e)) { OptimizerOptimal(e); }
\end{lstlisting}
}}
\subsection{Lights Prompt}
\textbf{Coq.}
{\it
In Coq:
(1) Write a datatype `light` for traffic lights with cases `Red`, `Yellow`, `Green`.
(2) Write a function `activation` which takes two lights, source and target, and returns a list of lights, the first element being the source and the last element being the target. If the source and target are not yellow and are distinct, then the returned list has a middle element of yellow.
(3) Write a helper `adjacent\_ok` that takes two lights, and checks that they are not one red and the other green.
(4) Write a helper `all\_adjacent\_ok` that takes a list of lights, and checks that all adjacent elements are `adjacent\_ok`.
(5) Write a lemma `check\_activation` to prove that forall source and target lights, a returned list never has adjacent elements that are distinct and red or green. The proposition should be `all\_adjacent\_ok (activation source target)`.

}
\textbf{Dafny.}
{\it
In Dafny:
(1) Write a datatype `light` for traffic lights with cases `Red`, `Yellow`, `Green`.
(2) Write a function `activation` which takes two lights, source and target, and returns a list of lights, the first element being the source and the last element being the target. If the source and target are not yellow and are distinct, then the returned list has a middle element of yellow.
(3) Write a helper `adjacent\_ok` that takes two lights, and checks that they are not one red and the other green.
(4) Write a helper `all\_adjacent\_ok` that takes a list of lights, and checks that all adjacent elements are `adjacent\_ok`.
(5) Write a lemma `check\_activation(source: light, target: light)` to prove that a returned list never has adjacent elements that are distinct and red or green. The `ensures` clause should be `all\_adjacent\_ok(activation(source, target))`.

}
\paragraph{Hints for Coq.}{\ \\
{\it
\#\#\# Hint: Start with `Require Import List. Import ListNotations.`
}}
\paragraph{Check lemma for Coq.}{\
{\begin{lstlisting}[language=dafny]
Lemma CHECK__check_activation: forall (source: light) (target: light),
    all_adjacent_ok(activation(source  target).
    Proof.
      intros.
      eapply check_activation; eauto.
    Qed.
\end{lstlisting}
}}
\paragraph{Check lemma for Dafny.}{\
{\begin{lstlisting}[language=dafny]
lemma CHECK__check_activation(source: light, target: light)
    ensures all_adjacent_ok(activation(source, target))
    {
      check_activation(source, target);
    }
\end{lstlisting}
}}
\subsection{BST Prompt}
\textbf{Coq.}
{\it
In Coq, (1) write an ADT for a tree of natural numbers. Call it `Tree`.
Then (2) write a predicate `IsBST` that checks whether a given tree is a binary search tree (BST).
Then (3) write a function `insert` that inserts an element into a binary search tree while preserving the BST property.
Then (4) write a predicate `Contains` that checks whether a given tree contains a given element.
Then (5) write a lemma `InsertContains` about the insert function that ensures that the tree resulting from inserting an element contains that element (without requiring nor ensuring the BST property).
Then (6) write another lemma `InsertPreservesBST` about the insert function that checks the BST property continues to hold after insertion. This lemma should take bounds on the BST, and require that the element to be inserted is within those bounds.

}
\textbf{Dafny.}
{\it
In Dafny, (1) write an ADT for a tree of natural numbers. Call it `Tree`.
Then (2) write a predicate `IsBST` that checks whether a given tree is a binary search tree (BST).
Then (3) write a function `insert` that inserts an element into a binary search tree while preserving the BST property.
Then (4) write a predicate `Contains` that checks whether a given tree contains a given element.
Then (5) write a lemma `InsertContains` about the insert function that ensures that the tree resulting from inserting an element contains that element (without requiring nor ensuring the BST property).
Then (6) write another lemma `InsertPreservesBST` about the insert function that checks the BST property continues to hold after insertion. This lemma should take bounds on the BST, and require that the element to be inserted is within those bounds.

}
\paragraph{Hints for Coq.}{\ \\
{\it
\#\#\# Hint: Start with `Require Import List. Import ListNotations.`\\
\\
\#\#\# Hint: Use Fixpoint instead of Definition for recursive functions.\\
\#\#\# Hint: Use `l` and `r` for variable names instead of `left` and `right` to avoid name clashes.
}}
\paragraph{Check lemma for Coq.}{\
{\begin{lstlisting}[language=dafny]
// (5) Lemma about the insert function that ensures the tree resulting
//     from inserting an element contains that element
Lemma CHECK_InsertContains: forall (t: Tree) (x: nat),
  Contains (insert t  x) x.
Proof.
  intros.
  eapply InsertContains; eauto.
Qed.

// (6) Lemma about the insert function that checks the BST property
//     continues to hold after insertion
lemma CHECK_InsertPreservesBST: forall (t: Tree) (x: nat) (min: nat) (max: nat),
  (IsBST t min max) -> min <= x <= max ->
  IsBST (insert t x) min max.
Proof.
    intros.
    eapply InsertPreservesBST; eauto.
Qed.
\end{lstlisting}
}}
\paragraph{Check lemma for Dafny.}{\
{\begin{lstlisting}[language=dafny]
// (5) Lemma about the insert function that ensures the tree resulting from
//     inserting an element contains that element
lemma CHECK_InsertContains(t: Tree, x: nat)
  ensures Contains(insert(t, x), x)
{
  InsertContains(t, x);
}

// (6) Lemma about the insert function that checks the BST property continues
//     to hold after insertion
lemma CHECK_InsertPreservesBST(t: Tree, x: nat, min: nat, max: nat)
  requires IsBST(t, min, max) && min <= x <= max
  ensures IsBST(insert(t, x), min, max)
{
    InsertPreservesBST(t, x, min, max);
}
\end{lstlisting}
}}
\subsection{Opt0 Prompt}
\textbf{Coq.}
{\it
In Coq, write an ADT for arithmetic expressions (called `Expr`) comprising constants, variables and binary additions. Then write an evaluator (called `Eval`) taking an expression and an environment (a function that takes a variable name and returns a number) and returning the number resulting from evaluation. Then write an optimizer (called `Optimize`) taking an expression and returning an expression with all additions by 0 removed. Then prove that the optimizer preserves the semantics as defined by the evaluation function. Do so by proving the lemma `OptimizePreservesSemantics: forall (e: Expr) (env: string -> nat), Eval(Optimize(e), env) = Eval(e, env)`.

}
\textbf{Dafny.}
{\it
In Dafny, write an ADT for arithmetic expressions (called `Expr`) comprising constants, variables and binary additions. Then write an evaluator (called `Eval`) taking an expression and an environment (a function that takes a variable name and returns a number) and returning the number resulting from evaluation. Then write an optimizer (called `Optimize`) taking an expression and returning an expression with all additions by 0 removed. Then prove that the optimizer preserves the semantics as defined by the evaluation function. Do so by proving the lemma `OptimizePreservesSemantics(e: Expr, env: string -> int) ensures Eval(Optimize(e), env) == Eval(e, env)`.

}
\paragraph{Hints for Coq.}{\ \\
{\it
\#\#\# Hint: In the optimizer, recursively optimize the sub-expressions.\\
\#\#\# Hint: You can import the `string` datatype with the line `Require Import Coq.Strings.String.`.\\
\#\#\# Hint: Use Fixpoint instead of Definition for recursive functions.\\
\#\#\# Hint: With tactics like `induction` and `destruct`, \_avoid\_ naming with `as` and let Coq pick the names for you. For example, use `induction e.` but \_not\_ `induction e as [...]`.\\
\\
\#\#\# Hint: For the proof, do `induction e.`. Do NOT name the hypotheses with `as`.\\
\#\#\# Hint: The simple cases are by `simpl. reflexivity.`.\\
\#\#\# Hint: The addition case is by `simpl. rewrite <- IHe1. rewrite <- IHe2. destruct (optimize e1); destruct (optimize e2); try destruct n; try destruct n0; eauto using PeanoNat.Nat.add\_0\_r.`.\\
\#\#\# Hint: You'll need `Require Import Arith`.
}}
\paragraph{Check lemma for Coq.}{\
{\begin{lstlisting}[language=dafny]
Lemma CHECK_OPS: forall (e: Expr) (env: string -> nat), Eval (Optimize e) env = Eval e env.
    Proof.
    intros.
    apply OptimizePreservesSemantics; eauto.
    Qed.
\end{lstlisting}
}}
\paragraph{Check lemma for Dafny.}{\
{\begin{lstlisting}[language=dafny]
lemma CHECK_OPS(e: Expr, env: string -> int)
    requires true
    ensures Eval(Optimize(e), env) == Eval(e, env)
{
    OptimizePreservesSemantics(e, env);
}
\end{lstlisting}
}}
\subsection{Factorial Prompt}
\textbf{Coq.}
{\it
In Coq, write a factorial function, called `fac`, and prove (in a lemma `FacPositive: forall (n: nat), fac n > 0`) that the factorial is always strictly positive.

}
\textbf{Dafny.}
{\it
In Dafny, write a factorial function, called `fac`, and prove (in a lemma called `FacPositive(n: nat)`) that the factorial is always strictly positive.

}
\paragraph{Hints for Coq.}{\ \\
{\it
\#\#\# Hint: Don't forget to import the Arith module.\\
\#\#\# Hint: use `Nat.lt\_0\_1` in the base case of the proof.\\
\#\#\# Hint: use `Nat.lt\_lt\_add\_r` in the inductive case of the proof.
}}
\paragraph{Check lemma for Coq.}{\
{\begin{lstlisting}[language=dafny]
Lemma CHECK_FacPositive: forall (n: nat), fac n > 0. Proof. intros. apply FacPositive; eauto. Qed.
\end{lstlisting}
}}
\paragraph{Check lemma for Dafny.}{\
{\begin{lstlisting}[language=dafny]
lemma CHECK_FacPositive(n: nat) ensures fac(n) > 0 { FacPositive(n); }
\end{lstlisting}
}}
\subsection{Food Prompt}
{\it
In Dafny:
(1) Write a datatype for `Food`: `Pasta` or `Pizza`. Each Pasta or Pizza has a list of toppings. Each `Topping` is one of: `tomato`, `cheese`, `olive`, `broccoli`, `mushroom`, `pepper`.
(2) Write a predicate `ok` that accepts any pizza with five toppings or fewer, and any pasta with two toppings or fewer.
(3) Write a lemma `ok3\_pizza` that proves that an accepted food with three or more toppings must be a pizza.

}
\paragraph{Hints for Dafny.}{\ \\
{\it
\#\#\# Hint: The length of a list or sequence `s` is `|s|`.
}}
\paragraph{Check lemma for Dafny.}{\
{\begin{lstlisting}[language=Dafny]
lemma CHECK_ok3_pizza(x: Food)
    requires ok(x)
    requires |x.toppings| >= 3
    ensures match x { case Pizza(_) => true case _ => false }
    {
      ok3_pizza(x);
    }
\end{lstlisting}
}}
\subsection{Reverse Prompt}
{\it
In Dafny:
(1) Write a function `reverse` that takes a list as input and reverses it.
(2) Then write a lemma `reverse\_permutes` that checks that for any list `l`, an element exists in `l` if and only if it exists in the result of calling `reverse` on `l`.
(3) Then write a lemma `reverse\_involutes` that checks that for any list `l`, calling `reverse` twice on `l` yields `l`.

}
\paragraph{Hints for Dafny.}{\ \\
{\it
\#\#\# Hint: The length of a list or sequence `s` is `|s|`.\\
\#\#\# Hint: Use a plain `function` to define `reverse`, not a `function method` or a `method`.\\

}}
\paragraph{Check lemma for Dafny.}{\
{\begin{lstlisting}[language=Dafny]
lemma CHECK__reverse_permutes(l: seq<int>)
    // TODO
    {
    }
    lemma CHECK__reverse_involutes(l: seq<int>)
    ensures reverse(reverse(l)) == l;
    {
      reverse_involutes(l);
    }
\end{lstlisting}
}}
\subsection{Days Prompt}
{\it
In Dafny:
(1) Write an ADT `Day` for the days of the week: `Sunday` to `Saturday`.
(2) Write a function `next\_biz\_day` that gives the next business day.
(3) Write a function`iter\_biz\_day(d: Day, n: nat): Day` that iterates the next business day function, for an arbitrary number n of business days.
(4) Write a lemma `iter5\_biz\_day\_idempotent` that ensures that starting with a business day, taking the next five business days is idempotent.

}
\paragraph{Check lemma for Dafny.}{\
{\begin{lstlisting}[language=Dafny]
lemma CHECK_iter5_biz_day_idempotent(d: Day)
    requires d != Saturday
    requires d != Sunday
    ensures iter_biz_day(d, 5) == d
    {
      iter5_biz_day_idempotent(d);
    }
\end{lstlisting}
}}

\section{Examples of Scoring Partial Programs}\label{sec:scoring-exs}
Partial program with a score of $0$:
\begin{lstlisting}[language=dafny]
datatype Expr = 
\end{lstlisting}

Partial program with a score of $+ 1$:
\begin{lstlisting}[language=dafny]
datatype Expr = 
    | Const(val: int)
\end{lstlisting}

Partial program with a score of $-1$:
\begin{lstlisting}[language=dafny]
datatype Expr = 
    | Const(val: int)
    | Var(name: string)
    | Add(e1: Expr, e2: Expr)

function Evaluate(e: Expr,
    env: string -> int): int
    reads env
{
    match e
    case Const(val) => val
    case Var(name) => env(name)
    case Add(e1, e2) =>
      Evaluate(e1, env) +
      Evaluate(e2, env)
}
\end{lstlisting}

The negative score is due to the \texttt{reads} clause, which shouldn't be there.
Unfortunately, we only confirm the error once the whole function is generated.

\section{Broader Impacts}\label{sec:broader-impacts}
The development of algorithms that allow generation of verified code using smaller models has notable broader impacts on both machine learning and society. We increase the efficiency per token in code language model usage, and allow for the usage of smaller models. This further reduces energy consumption and allows for a usage of cheaper hardware, thereby democratizing access to this technology. Our approach, which is asking models to prove their work is correct, and then immediately and externally checking whether the proof is correct, can mitigate some of the open issues with trusting LLMs.

\ifanon
\newpage
\section*{NeurIPS Paper Checklist}

\begin{enumerate}

\item {\bf Claims}
    \item[] Question: Do the main claims made in the abstract and introduction accurately reflect the paper's contributions and scope?
    \item[] Answer: \answerYes{} 
    \item[] Justification: The abstract and intro clearly state the key results directly.

\item {\bf Limitations}
    \item[] Question: Does the paper discuss the limitations of the work performed by the authors?
    \item[] Answer: \answerYes{} 
    \item[] Justification: Limitations is a subsection of our Discussion section.

\item {\bf Theory Assumptions and Proofs}
    \item[] Question: For each theoretical result, does the paper provide the full set of assumptions and a complete (and correct) proof?
    \item[] Answer: \answerNA{} 
    \item[] Justification: Our paper does not include theoretical results.

    \item {\bf Experimental Result Reproducibility}
    \item[] Question: Does the paper fully disclose all the information needed to reproduce the main experimental results of the paper to the extent that it affects the main claims and/or conclusions of the paper (regardless of whether the code and data are provided or not)?
    \item[] Answer: \answerYes{} 
    \item[] Justification: We discuss our algorithms in detail with pseudocode in section \ref{sec:mcts} and provide all used hyperparameters in Appendix \ref{sec:setup}.

\item {\bf Open access to data and code}
    \item[] Question: Does the paper provide open access to the data and code, with sufficient instructions to faithfully reproduce the main experimental results, as described in supplemental material?
    \item[] Answer: \answerYes{} 
    \item[] Justification: We provide our code, including instructions on how to run it and reproduce our experimental results.

\item {\bf Experimental Setting/Details}
    \item[] Question: Does the paper specify all the training and test details (e.g., data splits, hyperparameters, how they were chosen, type of optimizer, etc.) necessary to understand the results?
    \item[] Answer: \answerYes{}
    \item[] Justification: Yes, we specify all hyperparameters, algorithms, and and used models.

\item {\bf Experiment Statistical Significance}
    \item[] Question: Does the paper report error bars suitably and correctly defined or other appropriate information about the statistical significance of the experiments?
    \item[] Answer: \answerYes{} 
    \item[] Justification: The error bars report Wilson intervals as described in \ref{sec:T}.

\item {\bf Experiments Compute Resources}
    \item[] Question: For each experiment, does the paper provide sufficient information on the computer resources (type of compute workers, memory, time of execution) needed to reproduce the experiments?
    \item[] Answer: \answerYes{} 
    \item[] Justification: Experiments are explicitly reported in terms of token counts which can be directly converted to compute requirements on your hardware. We use an internal cluster with A100 and H100 GPUs.
    
\item {\bf Code Of Ethics}
    \item[] Question: Does the research conducted in the paper conform, in every respect, with the NeurIPS Code of Ethics \url{https://neurips.cc/public/EthicsGuidelines}?
    \item[] Answer: \answerYes{} 
    \item[] Justification: The paper conforms with the code of ethics.

\item {\bf Broader Impacts}
    \item[] Question: Does the paper discuss both potential positive societal impacts and negative societal impacts of the work performed?
    \item[] Answer: \answerYes{} 
    \item[] Justification: We discuss broader impacts in Appendix \ref{sec:broader-impacts}.
    
\item {\bf Safeguards}
    \item[] Question: Does the paper describe safeguards that have been put in place for responsible release of data or models that have a high risk for misuse (e.g., pretrained language models, image generators, or scraped datasets)?
    \item[] Answer: \answerNA{}
    \item[] Justification: This paper is about optimizing results from existing models, and does not introduce new models. Hence, we believe our paper poses no such risks.

\item {\bf Licenses for existing assets}
    \item[] Question: Are the creators or original owners of assets (e.g., code, data, models), used in the paper, properly credited and are the license and terms of use explicitly mentioned and properly respected?
    \item[] Answer: \answerYes{} 
    \item[] Justification: The model we use is correctly cited in section \ref{sec:base-model}.

\item {\bf New Assets}
    \item[] Question: Are new assets introduced in the paper well documented and is the documentation provided alongside the assets?
    \item[] Answer: \answerNA{} 
    \item[] Justification: The paper does not release new assets.

\item {\bf Crowdsourcing and Research with Human Subjects}
    \item[] Question: For crowdsourcing experiments and research with human subjects, does the paper include the full text of instructions given to participants and screenshots, if applicable, as well as details about compensation (if any)? 
    \item[] Answer: \answerNA{} 
    \item[] Justification: Our paper does not involve crowdsourcing nor research with human subjects.

\item {\bf Institutional Review Board (IRB) Approvals or Equivalent for Research with Human Subjects}
    \item[] Question: Does the paper describe potential risks incurred by study participants, whether such risks were disclosed to the subjects, and whether Institutional Review Board (IRB) approvals (or an equivalent approval/review based on the requirements of your country or institution) were obtained?
    \item[] Answer: \answerNA{}
    \item[] Justification: Our paper does not involve crowdsourcing nor research with human subjects.

\end{enumerate}
\fi

\end{document}